\documentclass[twocolumn,showpacs,preprintnumbers,amsmath,amssymb]{revtex4}
\usepackage{epsfig}

\begin{document}

\def\bea{\begin{eqnarray}}
\def\eea{\end{eqnarray}}
\def\a{\alpha}
\def\p{\partial} 
\def\nn{\nonumber}
\def\r{\rho}
\def\xb{\bar{x}}
\def\vb{\bar{v}}
\def\fb{\bar{f}}
\def\lab{\bar{\lambda}}
\def\la{\langle}
\def\ra{\rangle}
\def\f{\frac}
\def\o{\omega}
\def\P{\mathcal{P}}
\def\d{\delta}

\draft

\title{Reexamination of experimental tests of the fluctuation theorem}

\author{Onuttom Narayan$^1$ and Abhishek Dhar$^{1,2}$}
\affiliation{
$^1$Department of Physics, University of California, Santa Cruz, CA 95064\\
$^2$ Raman Research Institute, Bangalore 560080.}
\date{\today}

\begin{abstract}
The Fluctuation Theorem and the Jarzynski equality are examined in the
light of recent experimental tests. For a particle dragged through a
solvent, it is shown that $Q$, the heat exchanged with the reservoir,
does not obey the Jarzynski equality due to slowly decaying tails in
its distribution.  For molecular stretching experiments, substantial
corrections to the Jarzynski equality can result from not measuring the
force at the end of the molecule that is moved. We also present a proof
of the Fluctuation Theorem for Langevin dynamics that is considerably
simpler than the standard proof.  
\end{abstract}

\pacs{05.70.Ln, 82.20.Wt}

\maketitle

\section{Introduction}
Understanding how entropy behaves is the fundamental issue in statistical
mechanics.  Equilibrium statistical mechanics provides a powerful
framework to examine a broad variety of systems, but is by its nature
restricted to systems in equilibrium.  Recently, a remarkable Fluctuation
Theorem was proved~\cite{Gallavotti} for entropy growth or decay for
systems in a non-equilibrium steady state where the microscopic dynamics
are time reversal invariant. This theorem states that in a time interval
$\tau,$ the probability $p(S)$ of an entropy $S$ being generated satisfies
the condition
\begin{equation}
\lim_{\tau\rightarrow\infty}\ln{{p(S)}\over{p(-S)}} = S/k_B
\label{SSFT}
\end{equation}
where $k_B$ is Boltzmann's constant. The theorem was motivated by
numerical results on a shear stress model~\cite{Evans}. In Eq.(\ref{SSFT})
entropy is defined in dynamical systems terms, through phase space
contraction and expansion, rather than thermodynamically.

When instead of being in a non-equilibrium
steady state, the system starts in thermal equilibrium and then subject 
to a time-independent perturbation, a stronger
result can be proved~\cite{Evans2}:
\begin{equation}
\ln{{p(S)}\over{p(-S)}} = S/k_B
\label{TFT}
\end{equation}
valid irrespective of the length of the observation time interval
$\tau.$ Eqs.(\ref{SSFT}) and (\ref{TFT}) are referred to as the Steady
State Fluctuation Theorem (SSFT) and the Transient Fluctuation Theorem
(TFT) respectively. A simple example illustrating the need for the
$\tau\rightarrow\infty$ limit in the first but not the second is
given in Ref.~\cite{Zon-Cohen}. The fluctuation theorem was first
proved for thermostatted Hamiltonian systems, and later for systems
undergoing Langevin dynamics~\cite{Kurchan}. Apart from numerical
realizations~\cite{Evans,numeric}, experimental tests of the
fluctuation theorem have also been performed~\cite{Wang,Menon}.

A related result was proved by Jarzynski~\cite{Jarzynski}, relating
the change in free energy of a system that starts in equilibrium at a
temperature $T,$ when it is perturbed externally by making its Hamiltonian
time dependent. If the Hamiltonian starts and stops changing at time $t=0$
and $t=\tau$ respectively, and $W_J$ is the generalized work defined as
$\int_0^\tau dt \partial_t H(t),$ then the Jarzynski equality states that
\begin{equation}
\langle\exp[-\beta W_J]\rangle = \exp[-\beta \Delta F]
\label{Jarzynski}
\end{equation}
where $\beta = 1/k_B T.$ The change in the free energy of the system between
equilibrium at temperature $T$ with Hamiltonian $H(0)$ and 
equilibrium at the same temperature with Hamiltonian $H(\tau)$ is 
denoted by $\Delta F.$ The Jarzynski equality relates the change in free 
energy, an equilibrium concept, to an average of nonequilibrium measurements.
It has been used in experiments where a molecule is stretched at a finite
rate (i.e. not adiabatically) to obtain the free energy change due to 
stretching~\cite{Liphardt}. The connection between the Jarzynski equality 
and the TFT was shown by Crooks~\cite{Crooks} who showed that if an externally
driven system starting in thermal equilibrium is compared to the same system 
with the external driving time-reversed, and there is a quantity $\omega$ that 
satisfies 
\begin{equation}
{{p_+(\omega)}\over{p_-(-\omega)}} = \exp[\omega]
\label{crooks1}
\end{equation}
where the $\pm$ subscript refer to the original and the time-reversed process
respectively, then 
\begin{equation}
\langle\exp[-\omega]\rangle_{+,-} = 1.
\label{crookseq}
\end{equation}
He then showed that $S/k_B$ is expected to satisfy Eq.(\ref{crooks1})
for $\omega,$ and related $\omega$ to $W_J$ on thermodynamic grounds
yielding Eq.(\ref{Jarzynski}). When the original and time-reversed driving
are equivalent, as with a steady state perturbation, the subscripts in
Eq.(\ref{crooks1}) can be dropped, and Eq.(\ref{TFT}) results.

Despite the dramatic progress in the field, some uncertain points
remain. The relationship between the actual physical work done on a system
and the generalized work $W_J$ can be tricky~\cite{Hummer,Schurr}. Since
Crooks' connection~\cite{Crooks} between the TFT and the Jarzynski
equality uses thermodynamic arguments, it deals with the physical work
$W$ rather than $W_J.$ Questions have also been raised about possible
alternative definitions of entropy generated~\cite{Mazonka, Zon}, and
whether the equations above would be satisfied. There are also issues
about the experiments, and whether they measure what they should. In
these paper, we seek to clarify some of these points.

The organization of the rest of this paper and its main results
are as follows. Section II discusses the connection between $W_J$
and the physical work $W$ in various different cases, showing that
Eq.(\ref{Jarzynski}) (with the free energy appropriately defined)
is always obtained.  Section III considers an alternative definition
of entropy fluctuations~\cite{Mazonka, Zon} in an experimental
context~\cite{Evans2}, and shows that Eq.(\ref{crookseq}) is not obeyed;
the violation of Eq.(\ref{crookseq}) grows worse as the measuring
time interval $\tau$ is increased. Section IV considers the molecular
stretching experiments~\cite{Liphardt} and whether they satisfy the
conditions required to invoke the Jarzynski equality, demonstrating
that in general there are substantial corrections. Finally, Section V
presents a very simple proof of the fluctuation theorem for Langevin
dynamics~\cite{Kurchan}, generalizes it when the dynamical equations
are linear, and discusses extensions of Eq.(\ref{Jarzynski}).

\section{Physical work and generalized work}
Eq.(\ref{Jarzynski}) is stated in terms of the generalized work $W_J,$
and was derived~\cite{Jarzynski} for a system at constant volume, in
which case $\Delta F$ is the change in Helmholtz free energy, $\Delta A$.
On the other hand, Crooks~\cite{Crooks} obtained the same result on
thermodynamic grounds, which therefore involves the actual work $W$
done by the external agent that changes the Hamiltonian. As pointed
out previously~\cite{Hummer,Schurr}, the two are not always equal. For
example, if a time dependent force $f(t)$ is applied to a polymer
chain, it is equivalent to a time dependent term in the Hamiltonian
$-f(t) x.$ The work done along a trajectory $\int dt f(t) \dot x(t)$
differs from $W_J = - \int x df(t)$ by $f(\tau)x(\tau) - f(0) x(0).$
However, $W_J$ is not {\it always\/} different from the physical work.
If the same polymer chain is stretched by holding its end at a fixed
location and moving this location, $\partial_t H = \dot x\partial_x H,$
and the force exerted on the end by the polymer is equal to $-\partial_x
H,$ so that the work done by the external agent in moving the end is
equal to $W_J$~\cite{foot_endpt}.  We distinguish, therefore, between
experiments where the variation of the Hamiltonian can be expressed as
the change of a coordinate, and those where it can be expressed as the
change of a force. In the first case, there is no difference between the
actual work and $W_J$~\cite{foot1}. Since in the latter case there is an
apparent discrepancy between the result of Crooks and that of Jarzynski,
it is worth clarifying the situation. For completeness, we also allow for
the possibility where the unperturbed system is held at fixed pressure
(or similar intensive parameter) rather than fixed volume.

As a simple example, to make the differences between these cases clear,
we consider a gas in a container in thermal contact with a reservoir at a
temperature $T.$ The partition between the container and the reservoir is
either at a fixed location, or free to move to maintain equal pressures
in the container and the reservoir. Another wall of the container has a
movable piston, which either has a definite time-varying 
position or has a definite force applied to it.  When the reservoir
has fixed volume, and the piston's position is varied in a definite
manner, the Jarzynski equality is equivalent to $\langle \exp[-\beta W]
\rangle = \exp[-\beta\Delta A],$ and there is no difference between $W$
and $W_J$~\cite{footjosh}. We proceed to analyze the remaining two cases.

When the reservoir has a fixed pressure, and the piston's position is
varied in a definite manner, microscopic reversibility~\cite{footmenon}
in the dynamics ensures that the probability to follow a path $x(t)$ when
the piston is moved along a trajectory as compared to the probability
to move along the time reversed path if the piston is moved along the
opposite trajectory is given by
\begin{equation}
{{{\cal P}[x(+t)|\lambda(+t)]}\over{{\cal P}[\overline x(-t) | 
\overline\lambda(-t)]}} = \exp[-\beta Q\{x(+t), \lambda(+t)\}] 
\label{microscopic}
\end{equation}
where $Q\{x(+t), \lambda(+t)\}$ is the heat transferred to the system
from the bath along the forward trajectory, and $\lambda(t)$ denotes
the variation of the external (coordinate) parameter. The notation
is the same as in Ref.~\cite{Crooks}, but our result differs in the
absence of a $p\Delta V$ term in the exponential on the right hand side.
Eq.(\ref{microscopic}) is justified by noting that if a microstate $A$ of
the system and the reservoir together evolves with time into a microstate
$B,$ then the time reversed microstate $B^\prime$ must evolve into the
time reversed microstate $A^\prime$ when the variation of the external
parameter is reversed. The right hand side of Eq.(\ref{microscopic})
is then simply the ratio of the probability of finding the reservoir in
the appropriate microstate that would move the system along the forward
trajectory as compared to the backward one.  Using a constant pressure
ensemble for the reservoir, this ratio of probabilities is equal to
$\exp[\beta(U_f^R - U_i^R +p V_f^R - p V_i^R)],$ where $U^R_{i,f}$ and
$V^R_{i,f}$ are the energy and volume of the reservoir in the initial
and final state. (One can verify that, in the absence of any external
work, such a ratio of probabilities weights states of the system in
accordance with the constant pressure ensemble.) Since $\Delta U^R +
p\Delta V^R = -Q,$ Eq.(\ref{microscopic}) follows. 

Proceeding as in Crooks~\cite{Crooks} with the necessary changes, if 
the variable $\omega$ is defined as
\begin{equation} 
\omega = \ln p(x_i) - \ln p(x_f) - \beta Q 
\label{omega} 
\end{equation}
where $p_{i,f}$ are the probabilities of the initial and final states
of the system, then Eq.(\ref{crookseq}) is satisfied~\cite{foot_crooks}.
Using the fact that $\ln p_{i,f} = - \beta (U_{i,f} + p V_{i,f}) + \beta
G_{i,f},$ where $G$ is the Gibbs free energy of the system, we obtain
that $\omega = \beta (\Delta U + p\Delta V - Q - \Delta G) = \beta(W -
\Delta G),$ and so
\begin{equation} 
\langle\exp[-\beta W]\rangle = \exp[-\beta \Delta G].
\end{equation}
Thus the actual physical work is related to the change in the Gibbs free
energy. If the process is carried out adiabatically, $W$ is always equal
to $\Delta G.$

On the other hand, when the reservoir has a fixed volume but
the {\it force\/} on the piston is varied in a definite manner,
Eqs.(\ref{microscopic}), (\ref{omega}) and (\ref{crookseq}) still hold.
However, $\ln p_{i, f} = -\beta(U_{i, f} + F_{i, f} x_{i,f}) + \beta
G_{i,f},$ where $F$ and $x$ are the force exerted by the piston and
its coordinate.  $G = k_B T \ln \sum\exp[-\beta (U + F x)];$ if $F x$
is absorbed into $U$ as in~\cite{Jarzynski}, this would be considered
as the Helmholtz free energy. Now since $\Delta U = Q - W,$ where $W =
\int F dx,$ Eq.(\ref{crookseq}) yields
\begin{equation}
\langle\exp[\beta \int x dF]\rangle = \langle\exp[-\beta W_J]\rangle 
= \exp[-\beta\Delta G]
\end{equation}
in agreement with Jarzynski's result~\cite{Jarzynski}. 

It is also possible to consider the case when the reservoir is separated
from the system by a movable barrier (that equalizes pressures on both
sides) and the external force is applied to the same barrier. As expected,
in this case the appropriate free energy is obtained from $U + pV + Fx,$
while $W_J = - \int x dF.$ ($V$ and $x$ are not independent variables.)

To summarize, if a {\it coordinate} of a system is varied in a specific
way, the work done is related to the change in the appropriate free
energy, but if a {\it force\/} on a system is varied, the Legendre
transform of the work is related to the change in the appropriate
free energy. Although the thermodynamic approach starts with the work 
in all cases, the final result involves the generalized work, in 
complete agreement with~\cite{Jarzynski}.

To make matters clearer, we consider the experimentally relevant
case~\cite{Liphardt}, to which we shall return later in Section IV,
in which a molecule is stretched by placing its end in a parabolic
optical trap whose position is varied. If $x_0(t)$ is the center of
the trap and $x$ is the end of the molecule, the energy in the trap is
$k [x-x_0(t)]^2/2.$ If $H(x)$ is the internal energy of the molecule,
the Jarzynski equality relates the change in free energy with $H(\tau)
= H(x) + k[x - x_0(\tau)]^2/2$ to $W_J = -\int k [x - x_0(t)] dx_0.$
If one considers the trap and the molecule as a single combined system,
$W_J$ is the work done by the external agent moving the trap, and $H(t)$
is the appropriate Hamiltonian for the combined system. On the other hand,
if one is only interested in the molecule, then it would be desirable to
find a way to exclude the $k[x-x_0(\tau)]^2/2$ term when computing the
change in free energy. This latter option might seem more natural, and is
explored in~\cite{Hummer} and~\cite{Schurr}. However one could consider
an alternative experiment without an optical trap, with $x$ and $x_0$
the positions of the second-last and last sites on the molecule, and $x_0$
dragged along a definite path. The calculations would then be identical,
but the $k[x-x_0]^2/2$ term --- or an anharmonic generalization thereof
--- would be part of the stretching energy of the molecule, and $W_J$
would be the work done. No manipulations of the Jarzynski equality would
then be sought.

\section{Alternative definition of entropy fluctuations}
Wang et al~\cite{Wang} have performed experiments where a colloidal
particle in an optical trap is dragged in a solvent. The trap is
initially at rest, and then from time $t=0$ to $\tau$ is dragged at a
uniform velocity $v_0.$ The instantaneous force exerted by the optical
trap is $F_{\rm opt} = - k (x - x_0),$ where $x$ is the position of the
particle and $x_0$ is the position of the center of the trap. The work
done by the optical trap on the particle is then equal to
\begin{equation}
W(\tau) = - k \int_0^\tau dt v_0 (x - x_0).
\label{eq21}
\end{equation}
An integrated form of the fluctuation theorem
\begin{equation}
{{\Pr(W<0)}\over{\Pr(W>0)}} = \langle \exp[-\beta W]\rangle_{W>0}
\label{eq22}
\end{equation}
is verified~\cite{Wang}.

As discussed by Mazonka and Jarzynski~\cite{Mazonka,Zon}, the
entropy generated can be defined as $\Delta S = W/T$ or as $\Delta
S = - Q/T,$ where $Q$ is the heat absorbed by the particle from the
solvent, which acts as a heat reservoir.  The first definition, used
by Kurchan~\cite{Kurchan} and Crooks~\cite{Crooks}, is equivalent to
choosing~\cite{Crooks} the entropy of the particle in a certain state to
be $-\ln p,$ where $p$ is the probability of the state, and is justified
by recalling that the entropy of the particle over the entire canonical
ensemble is $\langle - \ln p\rangle.$ The second definition adopts the
convention that the entropy of the particle for a specified position and
velocity is zero, since it has no extra internal degrees of freedom, so
that entropy is only generated by heat being transferred to the solvent.
Both definitions are reasonable; to some extent, the entropy of a single
particle in a specific configuration is arbitrary.

With $\Delta S = W/T,$ it is possible to verify~\cite{Mazonka,Zon}
with Langevin dynamics for the particle that the TFT is satisfied,
in accordance with the experimental result~\cite{Wang}. A corollary
of this result is that the Jarzynski equality is satisfied, which
in this context is $\langle \exp[-\beta W]\rangle = 1.$ It is also
shown~\cite{Mazonka,Zon} that the SSFT is valid. The question was
raised~\cite{Mazonka} whether the fluctuation theorem --- either
transient or steady state --- is satisfied with the alternative choice
of $\Delta S = - Q/T.$ In Ref.~\cite{Zon}, it was shown that even in
the large time limit, if $Q/\tau$ is held constant, the fluctuation
theorem has to be modified for $Q$~\cite{comment}. In this section we
show that, $\langle\exp[\beta Q]\rangle$, analogous to the expression for $W$
in the Jarzynski equality, diverges as $\tau\rightarrow\infty,$ and also
for any finite $\tau$ in the commonly used limit when the mass of the
particle $m\rightarrow 0.$

Because the limit of zero particle mass is singular for $\langle\exp[\beta
Q]\rangle,$ we start with the Langevin equation for the particle
\begin{equation}
m \ddot x = - \lambda \dot x - k (x - x_0) + \eta(t)
\label{langevin}
\end{equation}
where $m$ is the mass of the particle, $\lambda$ is a viscous damping
coefficient, and $\eta(t)$ is the thermal noise satisfying $\langle
\eta(t_1)\eta(t_2)\rangle = 2\lambda k_B T \delta(t_1 - t_2).$ The center
of the optical trap is specified by $x_0= v_0 t \theta(t).$ 
The solution to Eq.(\ref{langevin}) can be written as
\begin{equation}
x(t) = \overline x(t) + \tilde x (t),
\label{separation}
\end{equation}
where
\begin{equation}
m\ddot{\overline x} = - \lambda \dot{\overline x} - k (\overline x - x_0)
\label{xbar}
\end{equation}
and
\begin{equation}
m\ddot{\tilde x} = - \lambda \dot{\tilde x} - k \tilde x + \eta.
\label{xprime}
\end{equation}
From Eq.(\ref{eq21}), the heat generation rate equal to
\begin{equation}
-dQ/dt = -k v_0 (x - x_0) - {1\over 2}{d\over{dt}}[k (x - x_0)^2 + 
m \dot x^2] = (\lambda \dot x - \eta)\dot x
\end{equation}
where Eq.(\ref{langevin}) is used to obtain the second form. Using
Eq.(\ref{separation}) and Eq.(\ref{xprime}) yields the result
\begin{equation}
-dQ/dt = \lambda \dot{\overline x}^2 + \dot{\overline x} 
[2\lambda \dot{\tilde x} - \eta] 
- {1\over 2} {d\over{dt}}[k \tilde x^2 + m \dot{\tilde x}^2].
\label{heateq}
\end{equation}
Sufficiently long after the optical trap starts to move,
$\dot{\overline x}\rightarrow v_0.$ Therefore, if Eq.(\ref{heateq})
is integrated over a sufficiently long time interval $[0,\tau],$, it 
is clear
that the three contributions to $-Q(\tau)$ are i) a deterministic term
that approaches $\lambda v_0^2 \tau$ ii) a stochastic term that is $\sim
O(\sqrt \tau)$, and iii) a random $\sim O(1)$ contribution from the total
derivative. Despite the relative smallness of the third term, it is 
crucial in proving that the fluctuation theorem is not satisfied by
$Q$ when $\tau\rightarrow\infty$ with $Q/\tau$ fixed~\cite{Zon,comment},
because it affects the tails of the distribution where the fluctuation
theorem is tested.

Even without detailed calculations, it is
possible to understand the singularity of $\langle\exp[\beta Q]\rangle$
qualitatively. Both singularities come
from the third term to the right hand side of
Eq.(\ref{heateq}). If $m\rightarrow 0,$ $\dot{\tilde x}$ is uncorrelated
from one instant to the next. Therefore $m[\dot{\tilde x}^2(0) -
\dot{\tilde x}^2(\tau)]/2$ decouples from the other contributions to
$-Q(\tau).$ Since $\dot {\tilde x}(\tau)$ is drawn from the Gaussian
distribution, $\sim \exp[-\beta m \dot {\tilde x}^2(\tau)/2],$ we see
that the integral over $\dot {\tilde x}(\tau)$ in $\langle\exp[\beta
Q]\rangle$ diverges.  Even when the zero mass limit is not taken, the same
argument applies to $m \dot{\tilde x}^2(\tau)/2$ and to $k \tilde
x^2(\tau)/2$ in the $t\rightarrow\infty$ limit. The full calculation is
shown in the appendix.

\section{Stretching experiments on polymers}
In experiments by Liphardt et al~\cite{Liphardt}, the Jarzynski equality
is put to use in non-equilibrium stretching measurements on single RNA
molecules, to obtain the free energy change as a result of stretching.
The RNA molecule has polystyrene beads attached to the two ends. One
bead (bead A) is placed in an optical trap, whose center is kept fixed
throughout the experiment. The bead at the other end (bead B) is held at
a definite time-dependent position using a piezoelectric actuator. The
force exerted on the molecule by the optical trap is measured. If the
origin of the coordinate system is taken to be at the center of the
optical trap, this force is related to the position of bead A by
$F = - k x.$ Therefore, by measuring the force, the (fluctuating)
position of bead A can be found. If $x_0(t) + L$ is the position of bead B,
where $L$ is the unstretched length of the molecule, the stretching is
equal to $z= x_0 - x.$ The integral
\begin{equation}
w = -\int F dz = k \int x d(x_0 - x)
\label{work}
\end{equation}
is evaluated over a large number of trials, and the Jarzynski identity
is used to obtain the change in the free energy of the molecule as a
result of stretching.

Clearly, if the experiment is performed adiabatically, $w$ will always be
equal to the reversible work required to stretch the molecule, i.e. the
change in the free energy.  However, in the irreversible case, there
are two differences between $w$ and the actual work done that raise a
question about the applicability of the Jarzynski identity. The first
is that the force exerted on bead A, not bead B, is measured. The
second is that the stretching is measured rather than the displacement
of bead B. These may seem innocuous changes; indeed, the second might
even be considered desirable, since it attempts to eliminate the change
in the potential energy of bead A, which is not of interest. In order to
see whether these changes are indeed harmless, we consider a toy model
for the system that is admittedly unrealistic: where the RNA molecule is
also treated as a harmonic spring. In view of the unfolding transition
seen in the molecule, this is obviously wrong. Our objective, however,
is to investigate whether the difference between $w$ and the actual work
done is potentially significant. In the same spirit, the damping and
thermal noise forces experienced along the RNA molecule are neglected,
and the mass of the beads is taken to be zero.

In this toy model, the equation for bead A is
\begin{equation}
\lambda \dot x = - k x + \kappa (v_0 t - x) + \eta
\label{toyeq}
\end{equation}
where, as before, $k$ is the force constant of the optical trap,
$\lambda$ is the viscous damping coefficient, and $\eta$ is the thermal
noise. $\kappa$ is the force constant for the (harmonic version of)
the RNA molecule. Bead B is assumed to be moved at constant speed $v_0$
over a time interval $[0,\tau].$ Proceeding as in the previous section,
defining $t_0= \lambda/(k + \kappa),$ Eq.(\ref{toyeq}) has the solution
\begin{equation}
x(t) = {{\kappa v_0 t}\over{\kappa + k}} - {{\kappa v_0 t_0}
\over{\kappa + k}}\Big(1- \exp[-t/t_0]\Big) + \tilde x
\label{toysoln}
\end{equation}
where
\begin{equation}
\tilde x(t) = \tilde x(0)\exp[-t/t_0] + \int_0^t dt_1
{{\eta(t_1)}\over\lambda}\exp[-(t-t_1)/t_0].
\label{toyfluc}
\end{equation}
Substituting this solution in Eq.(\ref{work}) yields
\begin{eqnarray}
dw/dt &=& \Bigg[{{k \kappa v_0}\over{k + \kappa}}
\{t - t_0 (1 - e^{-t/t_0})\} + k\tilde x\Bigg]\cr
  &  & \Bigg[{{ v_0}\over{k + \kappa}}\{k + \kappa e^{-t/t_0}\}
-  \dot{\tilde x}\Bigg].
\label{toywork}
\end{eqnarray}
If $v_0\rightarrow 0$ while keeping $v_0 \tau$ fixed,
i.e. the adiabatic limit is taken, the only term in $w(\tau)$
that survives comes from integrating the product of the first term in
each bracket. This is the change in free energy of the harmonic RNA
molecule. Accordingly, the dissipative work satisfies
\begin{eqnarray}
dw_D/dt &=& {{k\kappa v_0^2}\over{(k + \kappa)^2}}[\kappa t e^{-t/t_0}
- t_0(k + \kappa e^{-t/t_0})(1 - e^{-t/t_0})] \cr
&+& {{k v_0 \tilde x}\over{k + \kappa}}[k + \kappa e^{-t/t_0}] -
{{k\kappa v_0\dot{\tilde x}} \over {k + \kappa}}
[t - t_0(1 - e^{-t/t_0})] \cr
&-& {1\over 2} k d \tilde x^2/dt.
\end{eqnarray}
Comparing with Eq.(\ref{heateq}), this has the same structure
of a deterministic term, a Gaussian stochastic term and a total
derivative. However, if $\tau >>t_0,$ it can be seen that the integral of
the deterministic term is {\it negative,\/} so that $\langle \exp[-\beta
w_D(\tau)]\rangle$ cannot be equal to 1.

To solve for the dynamics of the system correctly, one would need to
modify Eq.(\ref{toyeq}) by using the full nonlinear force from the
stretching of the molecule, and including the damping and thermal noise
acting along the length of the molecule. This would be difficult, and not
amenable to an analytical treatment. Experimentally~\cite{Liphardt}, good
agreement is found between the free energy determined by non-equilibrium
and adiabatic stretching.  It is not clear whether this is fortuitous,
or can be justified: perhaps the fact that the non-equilibrium nature of
the experiments are only seen to be significant when the RNA molecule
goes through its unfolding transition might make $w$ as defined in
Eq.(\ref{work}) better behaved. For instance, the fact that the bead
in the optical trap moves backwards in the unfolding direction means
that the result $\langle w_D(\tau)\rangle < 0$ found (for $t>> t_0$) for
the toy model will not be the case for the real system, so deviations
from the Jarzynski equality are less dramatic.

The reliable method would be to modify the experiment and move the optical
trap while keeping the other end fixed. If the force and displacement
of the optical trap are measured, as in the previous section, the
Jarzynski equality can be safely invoked~\cite{foot3}. Even with the
modified experiment, the displacement of the optical trap rather than
the stretching of the molecule must be used in computing the work.
As discussed briefly in Section II, the Jarzynski equality will then
yield the change in the free energy of the optical trap and the molecule
together, i.e.
\begin{equation}
\langle e^{-\beta w_J(\tau)}\rangle =
{{\int dx(\tau) \exp[- \beta\{ A^{\rm in}(x(\tau))+ {k\over 2} (x(\tau) - v\tau)^2\}]}
\over
{\int dx(0)[-\beta \{A^{\rm in}(x(0)) + k x(0)^2/2\}]}}
\label{hummer1}
\end{equation}
where $A^{\rm in}$ is the internal free energy of stretching of the
molecule, and $x(0),x(\tau)$ are the starting and ending positions of
bead A relative to the initial center of the trap. (Bead B is fixed
throughout.) On the other hand, the quantity of interest is $A^{\rm in}.$
%
An elegant method to obtain $A^{\rm in}$ is due to Hummer and
Szabo~\cite{Hummer}, which relies on a stronger version of the Jarzynski
equality:
\begin{equation}
\langle e^{-\beta w_J(\tau)}\delta(x(\tau) - x)\rangle =
{{\exp[- \beta\{ A^{\rm in}(x)+ k (x - v\tau)^2/2\}]}
\over
{\int dx(0)[-\beta \{A^{\rm in}(x(0)) + k x(0)^2/2\}]}}
\label{hummer2}
\end{equation}
where the quantity averaged on the left hand side is $\exp[-\beta
w_J(\tau)]/\epsilon$ for experimental trials when $x< x_\tau < x +
\epsilon$ and zero otherwise, for small $\epsilon.$ Summing both sides
of Eq.(\ref{hummer2}) yields (\ref{hummer1}).  Eq.(\ref{hummer2})
can be proved by invoking the Feynman Kac theorem~\cite{Kac}.
The $k(x -
v\tau)^2/2$ term in Eq.(\ref{hummer2}) can easily be transferred to the
left hand side. Numerics with different values of $\tau,$ with $x$ chosen
to be close to its optimal value for each $\tau,$ were used~\cite{Hummer}
to obtain the free energy profile. Note that it is {\it not\/} correct to
choose a different $\tau$ for each experimental trial so that $x(\tau)$
is always equal to $x$ and seek to apply Eq.(\ref{hummer2}), i.e. to
perform an experiment of variable duration with a definite extension
of the molecule at the end of each trial. Such a protocol would yield
\begin{eqnarray}
\langle e^{-\beta w_J(x)}\rangle&\propto&
\int_0^{\tau_M} d\tau \exp[-\beta \{A^{\rm in}(x) + {k\over 2} (x - v \tau)^2\}]\cr
&\propto& [1 + {\rm erf}(x\sqrt{\beta k/2})]\exp[-\beta A^{\rm in}(x)]
\end{eqnarray}
if $\tau_M\rightarrow\infty$.

\section{Fluctuation theorems for systems with
  Langevin dynamics}

We consider a system of particles with generalized coordinates
$x=\{ x_l \}$ and velocities $v=\{ v_l=\dot { x_l} \}$ and described
by the Hamiltonian 
\bea
H(x,v)=\sum_l \f{m_l v_l^2}{2} +V(x)
\eea
where $V$ is an interaction potential. We assume that the
system is in contact with a heat reservoir, and its time evolution is
describable by Langevin dynamics. A set of external forces $\{f_l\}$
act on the particles and perform work on it. Thus the equations of
motion are given by: 
\bea
m_l \ddot{x}_l= -\f{\p V}{\p x_l}+f_l(t)-\lambda_l\dot{x}_l+\eta_l
\label{lang}
\eea
where $\eta_l$ is Gaussian white noise with the correlator $\la
\eta_l(t) \eta_m(t') \ra = 2 \lambda_l k_B T \delta (t-t') \delta_{lm}$.
For stochastic systems, the fluctuation theorem --- and therefore
the Jarzynski equality --- was proved by Kurchan~\cite{Kurchan}. We
provide a simpler proof, through Eq.(\ref{microscopic}).
For
discrete systems, evolving, for example, through Monte Carlo dynamics,
Eq.(\ref{microscopic}) has been proved~\cite{Crooks}. Here we give a 
proof for systems evolving through Langevin dynamics.

The principle of microscopic reversibility relates the probability of
a  particular path in phase space to the probability of the
time-reversed path. Consider
the evolution of the system from time $t=0$ 
to $t=\tau$ through a path specified by
$\{x(t),v(t), f(t)\}.$ 
The probability of this path
is given by:
\bea
{\mathcal{P}}_+ &=& {\mathcal{N}} \exp[- \sum_l\f{\beta}{4 \lambda_l } \int_0^\tau
  dt \eta_l^2(t)] \nn \\
&=& {\mathcal{N}} \exp[-\sum_l \f{\beta}{4 \lambda_l}\int_0^\tau  dt
( m_l \ddot{x}_l+\f{\p V}{\p x_l}-f_l(t)+\lambda_l \dot{x}_l)^2  ]\nn \\
\eea
where ${\mathcal{N}}$ is a normalization constant. 
Now consider the time-reversed path given by $\{
\bar{x}'(t),\bar{v}'(t),\bar{f}'(t) \}=  
\{ \bar{x}(\tau-t), -\bar{v}(\tau-t), \bar{f} (\tau-t) \}$. The
probability of this path is:
\bea
{\mathcal{P}}_-&=& {\mathcal{N}} \exp[-\sum_l \f{\beta}{4 \lambda_l} 
  \int_0^\tau\!\!\!  dt ( m_l \ddot{x'}_l+\f{\p V}{\p x'_l}-f'_l(t)+\lambda_l
  \dot{x}'_l)^2  ]  \nn \\
&=& {\mathcal{N}} \exp[-\sum_l \f{\beta}{4 \lambda_l} \int_0^\tau\!\!\!  dt
( m_l \ddot{x}_l+\f{\p V}{\p x_l}-f_l(t)-\lambda_l\dot{x}_l)^2  ]\nn\\
 \eea
Taking the ratio of the two probabilities leads to the principle of
microscopic reversibility
\bea
\f{\P_+}{\P_-} &=& \exp[-\beta\sum_l \int_0^\tau\!\!\!  dt
[ m_l \ddot{x}_l+\f{\p V}{\p x_l}-f_l(t)]\dot{x}_l]  \nn \\
&=& \exp[-\beta Q],
\label{prmicr}
\eea
where $Q$ is the amount of heat transferred from the heat bath to the
system. The identification of $Q$ as the heat transfer can be seen
either by noting that $Q=\sum_l \int_0^\tau [ m_l \ddot{x}_l+\f{\p V}{\p
    x_l}-f_l(t)]\dot{x}_l= \sum_l \int_0^\tau dt [-\lambda_l \dot{x}_l +\eta_l]
\dot{x}_l$ which is clearly the energy flow from the heat bath.
Note that in the absence of external forces the principle of
microscopic reversibility reduces to the usual detailed balance
principle which states that:
\bea
\f{P(x_f, v_f | x_i, v_i) }{P(x_i,
  -v_i | x_f, -v_f) }=e^{-\beta
  [H(x_f,v_f)-H(x_i,v_i)]}, 
\label{path}
\eea 
where $P(x, v | x', v')$ denotes the
probability of being at $(x, v)$ at time $t=\tau$ given
that it was at $(x', v')$ at time $t=0$.
This follows if we integrate Eq.~(\ref{prmicr}) over all paths
between $(x_i,v_i)$ and $(x_f,v_f)$.
From Eqs.(\ref{prmicr}) and (\ref{path}), the fluctuation theorem
and the Jarzynski equality can be obtained as in Crooks~\cite{Crooks}, 
discussed in Section II. In particular, when the external forces are
switched on at time $t=0,$ 
\begin{eqnarray}
&&\langle\exp[\beta\int_0^\tau x\cdot df]\rangle 
= \langle \exp[\beta f(\tau)\cdot x(\tau) - \beta\int_0^\tau dt f\cdot v]\rangle\cr
&=& {{\langle\exp[-\beta H(x(\tau), v(\tau)) + \beta f(\tau)\cdot x(\tau)]\rangle_{x(\tau), v(\tau)}}
\over{\langle\exp[-\beta H(x(0), v(0))]\rangle_{x(0), v(0)}}}.
\label{jarzlang}
\end{eqnarray}
As discussed after Eq.(\ref{hummer2}), Eq.(\ref{jarzlang})
is true even without $x(\tau)$ being averaged, so that $\exp[\beta f(\tau)\cdot x(\tau)]$
can be canceled on both sides and 
\begin{equation}
\langle \exp[-\beta \int_0^\tau dt f\cdot v]\rangle = 1.
\end{equation}

If one compares the system starting in thermal equilibrium with a force $f(t)$ switched
on for $0<t <\tau,$ with the same system but with a force $f(\tau-t),$ 
then the variable $\omega$ in Ref.~\cite{Crooks} is 
\bea
\omega &=& \ln \rho(x_i,v_i)-\ln \rho(x_f,v_f)-\beta Q \nn \\
&=&  \ln \rho(x_i,v_i)-\ln
\rho(x_f,v_f)\nn\\
&&-\beta[H(x_f,v_f)-H(x_i,v_i) -\int_0^\tau dt f \cdot v]\nn\\
&=&  \beta\int_0^\tau dt f\cdot v.
\eea
When the force that is switched on is time independent, then Eq.(\ref{crooks1}) 
reduces to the result of Ref.~\cite{Kurchan}. In fact, {\it even\/} when $f(t)$
depends on time, if the equations of motion Eq.(\ref{lang}) are linear, the 
distribution of $\omega =  \beta\int_0^\tau dt f\cdot v$ is Gaussian for both 
the original and the time reversed process, i.e. 
\begin{equation}
{{p_+(\omega)}\over{p_-(-\omega)}} \propto {{\exp[ - (\omega - m_+)^2/2\sigma_+]}
\over{\exp[ - (\omega + m_-)^2/2\sigma_-]}}.
\label{equation}
\end{equation}
Comparing to Eq.(\ref{crooks1}) we see that $\sigma_+ = \sigma_-$  and
$m_+^2/\sigma_+ = m_-^2/\sigma_-,$ i.e. $m_+ = m_-$ (since $m_\pm>0$). 
Further, $m_\pm/\sigma_\pm = {1\over 2}.$ One can thus drop the subscripts 
in Eq.(\ref{crooks1}) for linear equations of motion for arbitrary $f(t).$

More generally, when the interaction potential
depends on a set of externally controlled parameters $\mu(t)$ so
that we have $H=H(x,v,\mu)$. In this case we get
\bea
Q &=& \sum_l \int_0^\tau [ m_l \ddot{x}_l+\f{\p V(x,\mu)}{\p
    x_l}]\dot{x}_l  \nn \\ 
&=&  \int_0^\tau dt [ \f{d}{dt} ( \sum_l \f{m_l v_l^2}{2}+V) -
  \sum_l \f{\p     V}{\p \mu_l} \dot{\mu_l} ] \nn \\
&=& H(x_f,v_f,\mu_f) - H(x_i,v_i,\mu_i)-W_J
\eea

Now if we take $\rho=e^{-\beta [H(x,v,\mu)-F(\mu)]}$ where 
$F(\mu)=-(1/\beta) \ln (\int d x d v e^{-\beta H(x,v,\mu)})$
is a generalized free energy then we get the Jarzynski result:
\bea
\la e^{-\beta W_J} \ra = e^{-\beta [F(\mu_f)-F(\mu_i)]}
\eea

The principle of microscopic reversibility can be used to derive some
other interesting results that are related to the Jarzynski result and
have been derived using other approaches. We again consider the case
with a time 
dependent Hamiltonian in which case Eq.~(\ref{prmicr}) gives
$e^{-\beta W_J} \P_+=e^{-\beta
  [H(x_f,v_f,\mu_f)-H(x_i,v_i,\mu_i)]} \P_-$. Integrating
both sides over all paths between the fixed initial and final points
then gives: 
\bea
&&\la e^{-\beta W_J} \ra_{(x_i,v_i)}^{(x_f,v_f)} \nn \\ &=&
P_R(x_i,-v_i| x_f,-v_f) e^{-\beta
  [H(x_f,v_f,\mu_f)-H(x_i,v_i,\mu_i)]},  
\label{iden1}
\eea
where $P_R(x_i,-v_i| x_f,-v_f)$ is the transition probability
under the action of the time reversed Hamiltonian.
If we now integrate over all  initial states chosen from the canonical
ensemble, Eq.(\ref{hummer2}) results.

\section{Conclusion}
In this paper, we have analyzed recent experimental tests of the
Fluctuation Theorem and the Jarzynski equality. For a particle
dragged through a solvent~\cite{Evans2}, the heat absorbed, which
has been suggested as an alternative definition~\cite{Mazonka,Zon}
of the entropy generated, violates the Jarzynski equality.  In the
experiments on molecular stretching~\cite{Liphardt}, the fact that
the force is not measured at the end of the molecule that is dragged
can lead to substantial deviations from the Jarzynski equality, even
changing the sign of the average `work'.  We have also provided a proof
of the fluctuation theorem for systems governed by Langevin dynamics
that is much simpler than the standard proof~\cite{Kurchan}, and have
generalized it slightly for linear Langevin equations.  Finally, we
have also clarified the connection between the thermodynamic work used
in a thermodynamic derivation of the Jarzynski equality~\cite{Crooks}
and the generalized work in the original derivation~\cite{Jarzynski},
verifying that Eq.(\ref{Jarzynski}) is obtained in all cases.

\begin{acknowledgments}
It is a pleasure to acknowledge useful discussions with Jan Liphardt,
Joshua Deutsch and Narayanan Menon. We also thank Michael Schurr and
Narayanan Menon for providing us with copies of their papers prior
to release. AD acknowledges support from the NSF under grant DMR 0086287.
\end{acknowledgments}

\appendix

\section{Gaussian integral}
In this appendix, we evaluate the singularity in
$\langle\exp[\beta Q]\rangle$ discussed in Section III. Defining Gaussian
variables
\begin{eqnarray}
y_1 &=& \int_0^\tau dt \dot{\overline x}[2\lambda\dot{\tilde x}(t) 
- \eta(t)]\cr
y_2 &=& \sqrt m \dot{\tilde x}(\tau)\cr
y_3 &=& \sqrt m \dot{\tilde x}(0)\cr
y_4 &=& \sqrt k \tilde x(\tau)\cr
y_5 &=& \sqrt k \tilde x(0)
\end{eqnarray}
we have
\begin{equation}
\beta Q(\tau) = - \beta\lambda\int_0^\tau \dot{\overline x}^2 
- \beta y_1 + {1\over 2}\beta [y_2^2 - y_3^2 + y_4^2 - y_5^2].
\end{equation}

From Eq.(\ref{xbar}), it is straightforward to see that
\begin{equation}
\overline x(t) = v_0 t - {{\lambda v_0 }\over k} 
+ A_1 \exp[-\alpha_1 t] + A_2\exp[-\alpha_2 t]
\end{equation}
where the two exponentially decaying terms are the solutions to the
homogeneous equation for $x$ from Eq.(\ref{langevin}), satisfying
\begin{equation}
\alpha_1 \alpha_2 = {k\over m}, \qquad 
\alpha_1 + \alpha_2 = {\lambda\over m}.
\label{roots}
\end{equation}
With the initial condition
$\overline x(0) = \dot{\overline x}(0) = 0,$
\begin{equation}
A_{1,2} = {{v_0[1 - \lambda \alpha_{2,1}/k]}
\over{\alpha_{1,2} - \alpha_{2,1}}}.
\end{equation}
Similarly, from Eq.(\ref{xprime}), one obtains
\begin{eqnarray}
\tilde x(t) &=& {{\alpha_1 \tilde x(0) + \dot{\tilde x}(0)}
\over{\alpha_1 - \alpha_2}} e^{-\alpha_2 t} +
{{\alpha_2 \tilde x(0) + \dot{\tilde x}(0)}\over{\alpha_2 - \alpha_1}} 
e^{-\alpha_1 t} \cr
&+& \int_0^t dt^\prime {{\eta(t^\prime)}\over m} 
{{e^{-\alpha_1(t - t^\prime)} - 
e^{-\alpha_2(t - t^\prime)}}\over{\alpha_2 - \alpha_1}}.
\label{xprsol}
\end{eqnarray}

Using Eq.(\ref{xprsol}), it is possible to obtain the correlation between
the Gaussian variables $y_1\ldots y_5,$ as follows:
\begin{eqnarray}
\langle y_1^2\rangle &=& 2\lambda k_B T \int dt
\dot{\overline x}^2(t)\cr
\langle y_1 y_2\rangle &=& \langle y_1 y_4 \rangle = 0\cr
\langle y_1 y_3 \rangle &=& {{2\lambda k_B T}\over{\sqrt m}} 
\int dt \dot{\overline x}(t) {{\alpha_2 e^{-\alpha_2 t} - 
\alpha_1 e^{-\alpha_1 t}}\over{\alpha_2 - \alpha_1}}\cr
\langle y_1 y_5 \rangle &=& {{2\lambda k_B T\sqrt k}\over m} 
\int dt \dot{\overline x}(t) {{e^{-\alpha_2 t} - 
e^{-\alpha_1 t}}\over{\alpha_2 - \alpha_1}}\cr
\langle y_2^2\rangle &=& \langle y_3^2\rangle = 
\langle y_4^2\rangle = \langle y_5^2\rangle = k_B T\cr
\langle y_2 y_4\rangle &=& \langle y_3 y_5\rangle = 0\cr
\langle y_3 y_4\rangle &=& - \langle y_2 y_5\rangle = 
\sqrt{k\over m} k_B T {{e^{-\alpha_1 \tau} - e^{-\alpha_2 \tau}}
\over{\alpha_2 - \alpha_1}}\cr
\langle y_2 y_3\rangle &=& \sqrt{m\over k} 
{d\over{d\tau}}\langle y_3 y_4\rangle\cr
\langle y_4 y_5 \rangle &=& k_B T 
{{\alpha_2 e^{-\alpha_1 \tau} - \alpha_1 e^{-\alpha_2 \tau}}
\over {\alpha_2 - \alpha_1}}.
\label{corrcoeff}
\end{eqnarray}
In terms of the correlation matrix $M_{ij} = \langle y_i y_j\rangle,$
\begin{eqnarray}
& &\langle\exp[\beta Q]\rangle \propto \int dy_1 \ldots dy_5 
\exp[-\beta\lambda\int \dot{\overline x}^2 -\beta y_1 \cr
&+& {1\over 2}\beta (y_2^2 - y_3^2 + y_4^2 - y_5^2) 
- {1\over 2} \sum_{ij} y_i M^{-1}_{ij} y_j]
\label{gaussint}
\end{eqnarray}
where the normalization is obtained performing the same integral with all
terms in the exponential dropped except for the last. The integral
is formally straightforward: defining the matrix
\begin{equation}
N = M^{-1} + \beta{\rm diag}(0, -1, 1, -1, 1)
\label{defN}
\end{equation}
one obtains
\begin{equation}
\langle\exp[\beta Q]\rangle = \exp[-\beta\lambda \int \dot{\overline x}^2
+ {1\over 2}\beta^2 N^{-1}_{11}]/\sqrt{\det N \det M}.
\label{detform}
\end{equation}

To see the singularity clearly we first take the $m\rightarrow 0$
limit. From Eq.(\ref{roots}), one of $\alpha_{1,2}$ must diverge in
this limit; we choose this to be $\alpha_2.$ Dropping all terms with
$\exp[-\alpha_2 \tau]$ in the correlation matrix $M,$ and defining $\epsilon
= \alpha_1 (m/k)^{1/2} = (k/m)^{1/2}/\alpha_2,$
\begin{equation}
{M\over{k_B T}} = \left(
\begin{matrix}
B(\tau)          & 0      & \epsilon C(\tau)     & 0     & C(\tau)          \cr
0             & 1      & -\epsilon^2 E(\tau)  & 0     & -\epsilon E(\tau)\cr
\epsilon C(\tau) & -\epsilon^2 E(\tau) & 1     & \epsilon E(\tau) & 0       \cr
0             & 0      & \epsilon E(\tau)     & 1     & E(\tau)          \cr
C(\tau)          & -\epsilon E(\tau)   & 0       & E(\tau)          & 1     \cr
\end{matrix}
\right)
\label{presingular}
\end{equation}
where
\begin{eqnarray}
B(\tau) &=& 2\lambda \int dt \dot{\overline x}^2(t)\cr
C(\tau) &=& - {{2\lambda \sqrt k}\over{m (\alpha_2 - \alpha_1)}}\int dt
\dot{\overline x}(t)\exp[-\alpha_1 t]\cr
E(\tau) &=& {{\alpha_2}\over{\alpha_2 - \alpha_1}}\exp[-\alpha_1 \tau].
\end{eqnarray}
Notice that $B(\tau), C(\tau), E(\tau)$ are not singular as $m\rightarrow 0.$
Changing variables from $y_2, y_4$ to
$ (y_2 + \epsilon y_4)/(1 + \epsilon^2)^{1/2}$ and
$(y_4 - \epsilon y_2)/(1 + \epsilon^2)^{1/2},$ the matrix $M/k_B T$
changes to
\begin{equation}
\left(
\begin{matrix}
B(\tau)    &   0   &  \epsilon C(\tau)   & 0    & C(\tau)  \cr
0       &   1   &      0           & 0    & 0     \cr
\epsilon C(\tau) &  0  & 1 & \epsilon\sqrt{1 + \epsilon^2} E(\tau) & 0 \cr
0 & 0 &\epsilon\sqrt{1 + \epsilon^2} E(\tau) & 1 &\sqrt{1 + \epsilon^2} E(\tau)\cr
C(\tau)  &   0   &   0   & \sqrt{1 + \epsilon^2} E(\tau)  & 1\cr
\end{matrix}\right).
\label{singular}
\end{equation}
The second row and column decouple from everything else. From
Eq.(\ref{defN}), the second row and column of the matrix $N$ are
then zero. (The diagonal matrix in Eq.(\ref{defN}) is not affected by the
change of variables from Eq.(\ref{presingular}) to Eq.(\ref{singular}).) 
Taking the $m\rightarrow 0$ limit in all other
rows and columns of $M,$ it is possible to see that this is the only
singularity in this limit in Eq.(\ref{detform}).

When $m$ is very small but not zero, it is easy to see that the
offdiagonal terms $M_{2i} = M_{i2}$ are $O(\exp[-\alpha_2 \tau]),$
while $M_{22}$ is still $k_B T.$ Therefore $N_{2i} = N_{i2}$ are
$O(\exp[-\alpha_2 \tau])$ and $N_{22}$ is $O(\exp[-2 \alpha_2 \tau]).$
In Eq.(\ref{detform}), $\det M \det N$ is $O(\exp[-2 \alpha_2 \tau]),$
while $N^{-1}_{11}$ is regular as $m\rightarrow 0.$ Therefore
\begin{equation}
\langle \exp[\beta Q]\rangle \sim O (\exp[\alpha_2 \tau]) 
\sim O(\exp[\lambda \tau/m])
\end{equation}
where
$\alpha_2 = [\lambda +(\lambda^2 - 4 k m)^{1/2}]/(2m)\approx \lambda/m$
has been used.

Even when $m$ is not small, in the $\tau\rightarrow\infty$ limit,
$\exp[-\alpha_{1,2} \tau] \rightarrow 0.$ In Eqs.(\ref{corrcoeff}), in this
limit all cross-correlations are zero except for $\langle y_1 y_3\rangle$
and $\langle y_1 y_5\rangle.$ Thus the second and fourth rows and
columns of the matrix $N$ are zero in this limit. More accurately, since
$\exp[-\alpha_2 \tau]<< \exp[-\alpha_1 \tau]$ one can proceed as above (without
assuming that $\epsilon$ is small), and obtain that
$N_{2i}=N_{i2} = O(\exp[-\alpha_2 \tau])$ and
$N_{22} = O(\exp[-2\alpha_2 \tau])$ as before. In addition, in the remaining
$4\times 4$ submatrix formed by eliminating the second row and column,
$N_{4i}=N_{i4} = O(\exp[-\alpha_1 \tau])$ and
$N_{44} = O(\exp[-2\alpha_1 \tau]).$ Therefore
\begin{equation} 
\langle \exp[\beta Q]\rangle \sim O (\exp[(\alpha_2 +\alpha_1) \tau]) 
\sim O(\exp[\lambda t/m]).  
\end{equation}
This has the same form as the previous equation, though the singularity in
the two cases come from the $m\rightarrow 0$ and the $\tau\rightarrow\infty$
limit respectively.


\begin{references}
\bibitem{Gallavotti} G. Gallavotti and E.G.D. Cohen, Phys. Rev. Lett. {\bf 74},
2694 (1995).

\bibitem{Evans} D.J. Evans, E.G.D. Cohen and G.P. Morriss, Phys. Rev. Lett.
{\bf 71} 2401 (1993). 

\bibitem{Evans2} D.J. Evans and D.J. Searles, Phys. Rev. E{\bf 50}, 1645 (1994).

\bibitem{Zon-Cohen} R. van Zon and E.G.D. Cohen, cond-mat/0212311.

\bibitem{Kurchan} J. Kurchan, J. Phys. A {\bf 31}, 3719 (1998).

\bibitem{numeric} F. Bonetto, G. Gallavotti and P.L. Garrido, Physica D {\bf 105},
226 (1997); S. Aumaitre, S. Fauve, S. McNamara and P. Poggi, Eur. Phys. J. B{\bf 19},
449 (2001); D.J. Searles and D.J. Evans, J. Chem. Phys. {\bf 113}, 3503 (2000);
D.J. Evans, D.J. Searles and E. Mittag, Phys. Rev. E{\bf 65}, 051105 (2001); 
G. Ayton, D.J. Evans and D.J. Searles, J. Chem Phys. {\bf 115}, 2033 (2001).

\bibitem{Wang} G.M. Wang, E.M. Sevick, E. Mittag, D.J. Searles and D.J. Evans,
Phys. Rev. Lett. {\bf 89}, 050601 (2002).

\bibitem{Menon} S. Ciliberto and C. Laroche, J. Phys. IV {\bf 8}, 215 (1998);
K. Feitosa and N. Menon, preprint.

\bibitem{Jarzynski} C. Jarzynski, Phys. Rev. Lett. {\bf 78}, 2690 (1997).

\bibitem{Liphardt} J. Liphardt, S. Dumont, S.B. Smith, I. Tinoco and C. Bustamante,
Science {\bf 296}, 1832 (2002).

\bibitem{Crooks}  g. Crooks, Phye. Rev. E{\bf 60}, 2721 (1999).

\bibitem{Hummer} G. Hummer and A. Sazbo, Proc. Nat. Acad. Sci. {\bf 98}, 3658 (2001).

\bibitem{Schurr} M. Schurr and B.S. Fujimoto, preprint.

\bibitem{Mazonka} O. Mazonka and C. Jarzynski, cond-mat/9912121.

\bibitem{Zon} R. van Zon and E.G.D. Cohen, Phys. Rev. Lett. (in 
press) and cond-mat/0305147.

\bibitem{foot_endpt} 
Strictly speaking, in the case of Langevin dynamics, the force exerted
on the end of the molecule by the external agent differs from the
stretching force by the viscous drag and thermal noise terms acting on
the end. However, recognizing that these terms ultimately arise from
Hamiltonian interactions with the reservoir, $W$ is equal to $W_J.$

\bibitem{foot1}
The imposition of a specified coordinate can be done through a hard
constraint rather than a potential. For instance, in the case of a
gas confined to a container, the Hamiltonian does not change when
the container expands, but the range over which the coordinates can
vary does. However, this can be understood as the singular limit of a
potential that is zero inside the container and grows extremely rapidly
outside the container. One can verify that the force exerted on the wall
is equal to the derivative of the wall-system potential (and therefore
the Hamiltonian) with respect to the wall coordinate, so that $W=W_J.$

\bibitem{footjosh}
Even in this case, $\Delta A$ is not the change in the Helmholtz free
energy of the system and the reservoir together, but the change that {\it
would\/} have occurred if their temperature had been held constant while
the system Hamiltonian was varied.  (Under such a hypothetical change,
the free energy of the reservoir is unchanged, so that the change in free
energy of the system and the reservoir together is equal to the change
in free energy of the system.) In reality, the work done by the external
agent changes the temperature slightly. Though for a sufficiently large
reservoir this temperature change is negligible, it still leads to a
finite change in the free energy of the reservoir: $\Delta A(res) = -
S(res) \Delta T =  [S(res)/C_v(res)]\Delta Q.$ We thank Joshua Deutsch
for discussions that helped clarify this point.

\bibitem{footmenon}
Microscopic reversibility is the requirement that when the velocities
of all degrees of freedom are reversed, {\it including\/} those of the
reservoir, the dynamics are reversed. This includes dissipative systems,
as long as there are no velocity-dependent forces. The improbability of
such a reversal is given by Eq.(\ref{microscopic}).

\bibitem{foot_crooks} 
Crooks considers $k_B\omega$ to be the entropy generated, a point that
is discussed further in the next Section of this paper. For the purpose
of deriving the Jarzynski equality, it does not matter what $\omega$ is.

\bibitem{comment} In an earlier version of this paper, it was claimed 
that the TFT for $Q$ is valid in the large time limit for  this experiment. 
The error in this was pointed out by R. van Zon and E.G.D. Cohen in 
cond-mat/0307297.

\bibitem{foot3} 
One might be tempted to modify Eq.(\ref{work}) as $dw = k x d x_0$ and
argue that this is equivalent to moving the optical trap from Galilean
invariance.  However, this neglects the damping from the solvent, and can
(with our toy model) be verified to be incorrect.

\bibitem{Kac} Z. Schuss, {\it Theory and applications of Stochastic
  Differential Equations} (Wiley, New York, 1980).

\end{references}
\end{document}